\documentstyle[twoside,fleqn,espcrc2]{article}
\newcommand{\AmS}{{\protect\the\textfont2
  A\kern-.1667em\lower.5ex\hbox{M}\kern-.125emS}}

\title{Axial-vector currents and $\tau$ mesonic decays}

\author{Bing An Li\address{Department of Physics and Astronomy,
        Univ. of Kentucky, \\
        Lexington, Kentucky}%
        \thanks{Work partially supported by the Department of Energy under
  grant DE-91ER75661.}}

\begin{document}

\begin{abstract}
A general expression of the axial-vector current is presented,
in which
both the effects of the chiral symmetry breaking and the spontaneous
chiral symmetry breaking are included. A new resonance formula of the
axial-vector meson is derived and in the limit of
$q^{2}\rightarrow 0$
this formula doesn't go back to the ``chiral limit``. The studies
show that the dominance of the axial-vector meson is associated with
the satisfaction of PCAC.
The dominance of pion exchange is companied by
the strong anomaly of PCAC.
\end{abstract}

\maketitle

\section{INTRODUCTION}
The $\tau$ mesonic
decays have been studied by many authors[1].
Chiral symmetry play an
important role in studying $\tau$ mesonic decays.
In the theory of $\tau$ mesonic decays there are three parts:
vector currents; axial-vector currents;  meson vertices.
The vector currents are treated by VMD.
In the chiral limit, we have
\begin{eqnarray}
\lefteqn{{\cal L}^{V}={g_{W}\over 4f_{\rho}}cos\theta_{C}
\{-{1\over 2}
(\partial_{\mu}A^{i}_{\nu}-\partial_{\nu}
A^{i}_{\mu})
(\partial_{\mu}\rho^{i}_
{\nu}-\partial_{\nu}\rho^{i}_{\mu})}\nonumber \\
&&+A^{i}_{\mu}j^{i\mu}\},
\end{eqnarray}
$j^{i}_{\mu}$ is derived by the substitution
\begin{equation}
\rho^{i}_{\mu}\rightarrow {g_{W}\over4f_{\rho}}cos\theta_{C}
A^{i}_{\mu}
\end{equation}
in the vertices involving $\rho$ mesons.

There are three terms in the axial-vector current.
It is known for a long time that $a_{1}$ meson is the chiral partner
of the $\rho$ meson, therefore, there should be a term which is
similar with VMD, axial-vector meson dominance.
However, $a_{1}$ is much heavier than $\rho$ and
the mass difference is resulted in
spontaneous chiral symmetry breaking. The second term is related
to the mass difference of $a_{1}$ and $\rho$ meson.
The third term comes from
the coupling between the pion and W-boson. Combining
these three terms,
the general expression of the axial-vector current is obtained
\begin{eqnarray}
\lefteqn{{\cal L}^{A}=
-{g_{W}\over 4f_{a}}cos\theta_{C}
\{-{1\over 2}(\partial_{\mu}A^{i}_{\nu}
-\partial_{\nu}A^{i}_{\mu})}\nonumber \\
&&(\partial_{\mu}a^{i}_
{\nu}-\partial_{\nu}a^{i}_{\mu})
+A^{i\mu}j^{iW}_{\mu}\}\nonumber \\
&&-{g_{W}\over 4}cos\theta_{C}
\Delta m^{2}f_{a}A^{i}_{\mu}a^{i\mu}\nonumber \\
&&-{g_{W}\over4}cos\theta_{C}
f_{\pi}A^{i}_{\mu}\partial^{\mu}\pi^{i},
\end{eqnarray}
where $j^{iW}_{\mu}$ is determined by
\[a^{i}_{\mu}\rightarrow {g_{W}\over 4f_{a}}cos\theta A^{i}_{\mu}.\]
There are two parameters $f_{a}$ and $\Delta m^{2}$.
Using Weinberg's first sum rule and the mass difference between $\rho$
and $a_{1}$ meson, the two parameters are determined
\begin{equation}
f^{2}_{a}=f^{2}_{\rho}(1-{f^{2}_{\pi}f^{2}_{\rho}\over m^{2}_{\rho}})
{m^{2}_{a}\over m^{2}_{\rho}},\;\;\;
\Delta m^{2}=f^{2}_{\pi}(1-{f^{2}_{\pi}f^{2}_{\rho}\over
m^{2}_{\rho}})^{-1}.
\end{equation}

An effective chiral theory of pseudoscalar, vector, and axial-vector
mesons has been proposed[2], in which the meson fields are
simulated by quark operators.
For example,
\[\rho^{i}_{\mu}=-\frac{1}{g_{\rho}m^{2}_{\rho}}\bar{\psi}\tau_{i}\gamma
_{\mu}\psi.\]
The realization of quark operator forms of meson fields is done by
constructing a proper Lagrangian
\begin{eqnarray}
\lefteqn{{\cal L}=\bar{\psi}(x)(i\gamma\cdot\partial+\gamma\cdot v
+\gamma\cdot a\gamma_{5}
-mu(x))\psi(x)}\nonumber \\
&&-\bar{\psi}(x)M\psi(x)\nonumber \\
&&+{1\over 2}m^{2}_{0}(\rho^{\mu}_{i}\rho_{\mu i}+
\omega^{\mu}\omega_{\mu}+a^{\mu}_{i}a_{\mu i}+f^{\mu}f_{\mu})
\end{eqnarray}
where \(a_{\mu}=\tau_{i}a^{i}_{\mu}+f_{\mu}\),
\(v_{\mu}=\tau_{i}\rho^{i}_{\mu}+\omega_{\mu}\), and
\(u=exp{i\gamma_{5}(\tau_{i}\pi{i}+\eta)}\).
This theory has following features:
chiral symmetry;
spontaneous chiral symmetry breaking;
quark condensate;
Vector Meson Dominance(VMD) is a natural result;
Weinberg sum rule and
KSFR sum rule are satisfied;
Wess-Zumino-Witten Lagrangain is the leading term of the imaginary
part of the effective Lagrangian;
constituent quark mass is determined;
large $N_{C}$ expansion is revealed and
all loop diagrams of mesons are at higher
order;
two parameters of Chiral Perturbation Theory are determined in the
chiral limit;
there are five parameters: three current quark masses, a cut-off,
and a universal coupling constant which is chosen to be \(g=0.39\);
theoretical results of masses, decay widths, $\pi\pi$ scattering
agree with data well.
This theory is $QCD$ inspired, self-consistent, and
phenomenologically successful. The cut-off $\Lambda$ is determined
to be 1.6GeV which makes the theory suitable for the studies of
$\tau$ mesonic decays. In this theory they are determined:
\(\Gamma_{\rho}=142 MeV\) and \(m_{a}=1.20GeV\).

We have applied this theory to study $\tau$ mesonic decays[3].
The VMD is a natural result of this theory and the expression of the
axial-vector current is derived. The meson vertices of the $\tau$ decays
are found from the effective Lagrangaian of mesons[2,3]. Therefore,
this theory provides
a unified study for $\tau$ mesonic decays. There are
additional features in the studies of $\tau$ mesonic decays[3]:
\begin{enumerate}
\item The vertex VPP depends on momentum. For example,
$f_{\rho\pi\pi}$ derived from the effective Lagrangian[3]
is no longer a constant, but a function of $q^{2}$
\[f_{\rho\pi\pi}(q^{2})={2\over g}\{1+\frac{q^{2}}{2\pi^{2}f^{2}_{\pi}}
[(1-{2c\over g})^{2}-4\pi^{2}c^{2}]\},\]
where \(c={f^{2}_{\pi}\over2gm^{2}_{\rho}}\). The radius of the charged
pion is determined to be
\[r^{2}_{\pi}={6\over m^{2}_{\rho}}+{3\over\pi^{2}f^{2}_{\pi}}\{
(1-{2c\over g})^{2}-4\pi^{2}c^{2}\}\]
\[=0.447fm^{2}\]
which is in excellent agreement with data.
\item The vertex AVP depends on momentum.
\[{\cal L}^{a_{1}\rho\pi}=\epsilon_{ijk}\{Aa^{i}_{\mu}
\rho^{j\mu}\pi^{k}-Ba^{i}_{\mu}\rho^{j}_{\nu}\partial^{\mu\nu}\pi^{k}
\},\]
\[A={2\over f_{\pi}}gf_{a}\{{m^{2}_{a}\over g^{2}f^{2}_{a}}
-m^{2}_{\rho}\]
\[+p^{2}[{2c\over g}+{3\over4
\pi^{2}g^{2}}(1-{2c\over g})] \]
\[+q^{2}[{1\over 2\pi^{2}g^{2}}-
{2c\over g}-{3\over4\pi^{2}g^{2}}(1-{2c\over g})]\},\]
\[B=-{2\over f_{\pi}}gf_{a}{1\over2\pi^{2}g^{2}}(1-{2c\over g}),\]
where p and q are momentum of $\rho$ and $a_{1}$ respectively.
\item The resonance formula of the axial-vector meson takes a new
form
\[\frac{\Delta m^{2}f^{2}_{a}-m^{2}_{a}+i\sqrt{q^{2}}\Gamma_{a}(q^{2})}
{q^{2}-m^{2}_{a}+i\sqrt{q^{2}}\Gamma_{a}(q^{2})},\]
where $\Delta m^{2}f^{2}_{a}$ is related to the spontaneous chiral
symmetry breaking. In the limit of $q^{2}\rightarrow0$ this formula
doesn't go to the "chiral limit".
\item The normalization constants of the vector and the axial-vector
mesons are different.
\end{enumerate}
The tree diagrams are at the leading
order in large $N_{C}$ expansion. All the calculations are done at the
tree level.

\section{$a_{1}$ dominance in $\tau\rightarrow \pi\pi\pi\nu$}
The vertices $a_{1}\rho\pi$, $\rho\pi\pi$, $W\rho\pi$ contribute to
this decay mode. This theory explains the existence of
a $\rho$ resonance in the final state of this decay mode.
There is a cancelation between the diagrams. After taking
the cancelation into account, we obtain
\begin{eqnarray}
\lefteqn{<\rho^{0}\pi^{-}|\bar{\psi}\tau_{-}\gamma_{\mu}\gamma_{5}
\psi|0>
=\frac{i}{\sqrt{4\omega E}}(\frac{q_{\mu}q_{\nu}}{q^{2}}
-g_{\mu\nu})}\nonumber \\
&&(Ag_{\nu\lambda}+Bk_{\nu}k_{\lambda})\epsilon^{*\nu}
_{\sigma}\nonumber \\
&&\frac{g^{2}f_{a}m^{2}_{\rho}-if^{-1}_{a}\sqrt{q^{2}}\Gamma_{a}
(q^{2})}
{q^{2}-m^{2}_{a}+i\sqrt{q^{2}}\Gamma_{a}(q^{2})}
\end{eqnarray}
The $a_{1}$ dominance in $\tau\rightarrow \rho\pi\nu$
is revealed and the dominance is caused by
the cancelation which leads to the axial-vector current
conservation in the chiral limit.
The branching ratios are computed to be
\[B(\tau\rightarrow\pi^{+}\pi^{-}\pi^{-}\nu)=
B(\tau\rightarrow\pi^{-}\pi^{0}\pi^{0}\nu)=
6.3\%.\]
The decay width is determined to be
\[\Gamma_{a}=386MeV.\]
The ratio of the decay amplitudes of the $a_{1}$ meson, d/s,
is calculated to
be -0.098. These results are in reasonable agreement with the data.
\section{$a_{1}$ dominance in $\tau\rightarrow f_{1}\pi\nu$}
$f_{1}$ meson is the chiral partner of $\omega$ meson[2] and
the mass formula of $f_{1}$ meson is derived in Ref.[2]
\[(1-{1\over2\pi^{2}g^{2}})m^{2}_{f_{1}}=6m^{2}+m^{2}_{\omega},
\;\;\;m_{f_{1}}=1.21GeV.\]
The vertex of $f_{1}(1285)a_{1}\pi$
\[{\cal L}^{f_{1}a_{1}\pi}={1\over \pi^{2}f_{\pi}}{1\over g^{2}}
(1-{1\over2\pi^{2}g^{2}})^{-1}\varepsilon^{\mu\nu\alpha\beta}
f_{\mu}\partial_{\nu}\pi^{i}\partial_{\alpha}a^{i}_{\beta}\]
is from the Wess-Zumino-Witten anomaly and it
is used to calculate the decay rate of this decay mode.
The narrow width of the decay $f_{1}\rightarrow\rho\pi\pi$ is
revealed from this vertex[2].
$a_{1}$ is dominant in $\tau\rightarrow f_{1}\pi\nu$.
\[B(\tau\rightarrow f_{1}\pi\nu)=2.91\times10^{-4}.\]
The data is $(6.7\pm1.4\pm2.2)\times10^{-4}$(CLEO recent measurement).
\section{Effective Lagrangian of \(\Delta s=1\) weak interactions}
Using the $K^{*}$ fields to substitute the $\rho$ fields
in Eq.(1), the VMD of \(\Delta s=1\) is obtained.
In Ref.[2] the chiral partner of
$K^{*}(892)$ meson, the $K_{1}$ meson,
is coupled to
\[\bar{\psi}\lambda^{a}\gamma_{\mu}\gamma_{5}\psi. \]
The mass of this $K_{1}$ meson is derived as
\[(1-{1\over2\pi^{2}g^{2}})m^{2}_{K_{1}}=6m^{2}+m^{2}_{K^{*}},\]
\[m_{K_{1}}=1.32GeV.\]
It is lower than the mass of
$K_{1}(1400)$ and greater than $K_{1}(1270)$'s mass.
The widths of three decay modes
($K_{1}\rightarrow K^{*}\pi$, $K\rho$,
$K\omega$) are calculated[2]. It is found that $K^{*}\pi$ channel
is dominant, however, $B(K\rho)$ is about $11\%$. The data
shows that the branching ratio of
$K_{1}(1400)$ decaying into $K\rho$ is very small. Therefore,
the meson coupled to the quark axial-vector current is not a pure
$K_{1}(1400)$ state, instead, it is a mixture of the two $K_{1}$
mesons. It is $K_{a}$.
\[K_{a}=cos\theta K_{1}(1400)+sin\theta K_{1}(1270),\]
\[K_{b}=-sin\theta K_{1}(1400)+cos\theta K_{1}(1270). \]
In this theory
the amplitude of
$\tau\rightarrow K_{a}\nu$ is from the tree diagrams and at $O(N_{C})$
[2]. The production of $K_{b}$ in $\tau$
decay is through loop diagrams of mesons which is at $O(1)$
in large $N_{c}$ expansion[2]. This theory predicts a small
branching ratio for $K_{b}$ production in $\tau$ decays.
The axial-vector current of \(\Delta s=1\) is obtained by
using $K_{a}$ to substitute $a_{1}$ in Eq.(3).
\section{$\tau\rightarrow K^{*}\pi\nu$}
Both the vector and axial-vector currents contribute to this decay.
The vertex
$\pi K^{*}\bar{K}^{*}$ contributes to the vector part and
it is from the WZW anomaly[3].
The vector matrix element is obtained
\begin{eqnarray}
\lefteqn{<\bar{K}^{*0}\pi^{-}|\bar{\psi}\lambda_{+}\gamma_{\mu}\psi|0>=
\frac{-1}{\sqrt{4\omega E}}{1\over\sqrt{2}}\frac{N_{C}}
{\pi^{2}g^{2}f_{\pi}}}\nonumber \\
&&\frac{m^{2}_{K^{*}}-i\sqrt{q^{2}}
\Gamma_{K^{*}}
(q^{2})}{q^{2}-m^{2}_{
K^{*}}+i\sqrt{q^{2}}\Gamma_{K^{*}}(q^{2})}\varepsilon^{\mu\nu
\alpha\beta}k_{\nu}p_{\alpha}\epsilon^{*\sigma}_{\beta},
\end{eqnarray}
where p and k are momentum of $K^{*}$ and pion respectively,
\(q=k+p\).

The axial-vector matrix element is obtained by using the vertices:
$K_{a}K^{*}\pi$, $KK^{*}\pi$. In the
chiral limit, the expression of the matrix element of the axial-vector
current is derived
\begin{eqnarray}
\lefteqn{<\bar{K}^{*0}\pi^{-}|\bar{\psi}\lambda_{+}\gamma_{\mu}
\gamma_{5}\psi|0>=}\nonumber \\
&&\frac{i}{\sqrt{4\omega E}}
{1\over\sqrt{2}}({q_{\mu}q_{\nu}\over q^{2}}-g_{\mu\nu})
\epsilon^{*\lambda}_{\sigma}\nonumber \\
&&\{\frac{g^{2}f_{a}m^{2}_{K^{*}}
-iqf^{-1}_{a}\Gamma_{K_{1}(1400)}(q^{2})}{
q^{2}-m^{2}_{K_{1}(1400)}+iq\Gamma_{K_{1}(1400)}(q^{2})}\nonumber \\
&&cos\theta
(A_{K_{1}(1400)}(q^{2})_{K^{*}}g^{\nu\lambda}+B_{K_{1}(1400)}
k^{\nu}k^{\lambda}) \nonumber \\
&&+\frac{g^{2}f_{a}m^{2}_{K^{*}}
-iqf^{-1}_{a}\Gamma_{K_{1}(1270)}(q^{2})}{
q^{2}-m^{2}_{K_{1}(1270)}+iq\Gamma_{K_{1}(1270)}(q^{2})}\nonumber \\
&&sin\theta
(A_{K_{1}(1270)}(q^{2})_{K^{*}}g^{\nu\lambda}\nonumber \\
&&+B_{K_{1}(1270)}
k^{\nu}k^{\lambda})\}
\end{eqnarray}
The amplitudes $A_{K_{1}(1400)}$, $B_{K_{1}(1400)}$,
$A_{K_{1}(1270)}$, and $B_{K_{1}(1270)}$ are determined by the decay
widths of $K_{1}(1400)$ and $K_{1}(1270)$.
\begin{eqnarray}
K_{1}(1400)=cos\theta K_{a}-sin\theta K_{b},\nonumber \\
K_{1}(1270)=sin\theta K_{a}+cos\theta K_{b}.
\end{eqnarray}
The vertex of $K_{1}VP$ is found from Ref.[2]
\begin{eqnarray}
\lefteqn{{\cal L}^{K_{1}VP}=f_{abc}\{AK^{a}_{1\mu}V^{b\mu}P^{c}
}\nonumber \\
&&-BK^{a\mu}V^{b\nu}\partial_{\mu}\partial_{\nu}P^{c}\}.
\end{eqnarray}
The amplitude $A^{K^{*}}_{K_{a}}$ is
determined to be
\begin{eqnarray}
\lefteqn{A(q^{2})^{K^{*}}_{K_{a}}={2\over f_{\pi}}gf_{a}\{
{m^{2}_{K_{a}}\over g^{2}f^{2}_{a}}-m^{2}_{K^{*}}}\nonumber \\
&&+m^{2}_{K^{*}}
[{2c\over g}+{3\over4
\pi^{2}g^{2}}(1-{2c\over g})]\nonumber \\
&&+q^{2}[{1\over 2\pi^{2}g^{2}}-
{2c\over g}-{3\over4\pi^{2}g^{2}}(1-{2c\over g})]\}.
\end{eqnarray}
$B_{K_{a}}$ is the same as B. The amplitudes $A^{K^{*}}_{
K_{b}}$ and $B^{K^{*}}_{K_{b}}$ are unknown and we take them as
parameters.
Both $K_{1}(1400)$ and $K_{1}(1270)$ decay to $K\rho$ and
$K\omega$. Using the $SU(3)$ coefficients, for both $K_{1}$, it
is determined
\[B(K\omega)={1\over3}B(K\rho).\]
This relation agrees with data reasonably well.
For the $K\rho$ decay mode
$A^{\rho}_{K_{b}}$ and $B^{\rho}_{K_{b}}$ are other
two parameters. In the decays of the two $K_{1}$ mesons the
momentum of pion or kaon is low, therefore, the decay widths are
insensitive to the amplitude B. We take
\[B^{K^{*}}_{K_{b}}=B^{\rho}_{K_{b}}\equiv B_{b}.\]
The decay width of the $K_{1}$ meson is derived
\begin{eqnarray}
\lefteqn{\Gamma_{K_{1}}=\frac{k}{32\pi}\frac{1}{\sqrt{q^{2}}
m_{K_{1}}}\{(3+{k^{2}\over m^{2}_{v}})A^{2}(q^{2})}\nonumber \\
&&-A(q^{2})B(q^{2}+m^{2}_{v})\frac{k^{2}}{m^{2}_{v}}
+{q^{2}\over m^{2}_{v}}k^{4}B^{2}\},
\end{eqnarray}
where \(q^{2}=m^{2}_{K_{1}}\), \(v=K^{*}, \rho\),
k is the momentum of pion or kaon
\[k=\{{1\over4m^{2}_{K_{1}}}(m^{2}_{K_{1}}+m^{2}_{v}-m^{2}_{P})
^{2}-4m^{2}_{v}\}^{{1\over2}},\]
$m_{P}$ is the mass of pion or kaon.

we choose the parameters as
\begin{eqnarray}
\lefteqn{\theta=30^{0},\;\;\;A^{K^{*}}_{b}=-4.5GeV,\;\;\;
A^{\rho}_{b}=5.0GeV,}\nonumber \\
&&B_{b}=0.8GeV^{-1},
\end{eqnarray}
from which the decay widths are obtained
\begin{eqnarray}
\lefteqn{\Gamma(K_{1}(1400)\rightarrow K^{*}\pi)=159MeV,}\nonumber \\
&&\Gamma(K_{1}(1400)\rightarrow K\rho)=10.5MeV,\nonumber \\
&&\Gamma(K_{1}(1270)\rightarrow K^{*}\pi)=12.4MeV,\nonumber \\
&&\Gamma(K_{1}(1270)\rightarrow K\rho)=26.8MeV.
\end{eqnarray}

Using the two matrix elements(7,8), the distribution of the decay
rate is derived
\begin{eqnarray}
\lefteqn{{d\Gamma\over dq^{2}}(\tau^{-}\rightarrow
\bar{K}^{*0}\pi^{-}\nu)={G^{2}\over
(2\pi)^{3}}\frac{sin^{2}\theta_{C}}{128m^{3}_{\tau}q^{4}}
(m^{2}_{\tau}-q^{2})^{2}}\nonumber \\
&&(m^{2}_{\tau}+2q^{2})
\{(q^{2}+m^{2}_{K^{*}}
-m^{2}_{\pi})^{2}-4q^{2}m^{2}_{K^{*}}\}^{{1\over2}}\nonumber \\
&&\{\frac{6}{\pi^{4}g^{2}f^{2}_{\pi}}\frac{m^{4}_{K^{*}}+q^{2}
\Gamma^{2}_{K^{*}}(q^{2})}{(q^{2}-m^{2}_{K^{*}})^{2}+q^{2}
\Gamma^{2}_{K^{*}}(q^{2})}\nonumber \\
&&[(p\cdot q)^{2}-q^{2}m^{2}_{K^{*}}]\nonumber \\
&&+|A|^{2}[1+{1\over12m^{2}_{K^{*}}q^{2}}(q^{2}-m^{2}_{K^{*}})^{2}]
\nonumber \\
&&-(BA^{*}+B^{*}A)\frac{(q^{2}+m^{2}_{K^{*}})}{24m^{2}_{K^{*}}
q^{2}}(q^{2}-m^{2}
_{K^{*}})^{2}\nonumber \\
&&+\frac{|B|^{2}}{48m^{2}_{K^{*}}q^{2}}(q^{2}
-m^{2}_{K^{*}})^{4}\}.
\end{eqnarray}
where p is the momentum of $K^{*}$, $q^{2}$ is the invariant mass
squared of $K^{*}\pi$, and
\[A=\frac{g^{2}f_{a}m^{2}_{K^{*}}
-i\sqrt{q^{2}}f^{-1}_{a}\Gamma_{K_{1}
(1400)}}{q^{2}-m^{2}_{K_{1}(1400)}+i\sqrt{q^{2}}\Gamma_{K_{1}(1400)}}
cos\theta A^{K^{*}}_{K_{1}(1400)}\]
\[+\frac{g^{2}f_{a}m^{2}_{K^{*}}-i\sqrt{q^{2}}f^{-1}_{a}\Gamma_{K_{1}
(1270)}}{q^{2}-m^{2}_{K_{1}(1270)}+i\sqrt{q^{2}}\Gamma_{K_{1}(1270)}}
sin\theta A^{K^{*}}_{K_{1}(1270)},\]
\[B=\frac{g^{2}f_{a}m^{2}_{K^{*}}-i\sqrt{q^{2}}f^{-1}_{a}
\Gamma_{K_{1}
(1400)}}{q^{2}-m^{2}_{K_{1}(1400)}+i\sqrt{q^{2}}\Gamma_{K_{1}(1400)}}
cos\theta B^{K^{*}}_{K_{1}(1400)}\]
\[+\frac{g^{2}f_{a}m^{2}_{K^{*}}-i\sqrt{q^{2}}f^{-1}_{a}\Gamma_{K_{1}
(1270)}}{q^{2}-m^{2}_{K_{1}(1270)}+i\sqrt{q^{2}}\Gamma_{K_{1}(1270)}}
sin\theta B^{K^{*}}_{K_{1}(1270)},\]
where
\begin{eqnarray}
\lefteqn{A^{K^{*}}_{K_{1}(1400)}=cos\theta A^{K^{*}}_{a}
-sin\theta A^{K^{*}}_{b}}\nonumber \\
&&B^{K^{*}}_{K_{1}(1400)}=cos\theta B_{a}-sin\theta B_{b},\nonumber \\
&&A^{K^{*}}_{K_{1}(1270)}=sin\theta A^{K^{*}}_{a}
+cos\theta A^{K^{*}}_{b} \nonumber \\
&&B^{K^{*}}_{K_{1}(1270)}=sin\theta B_{a}+cos\theta B_{b}.
\end{eqnarray}
In the range of $q^{2}$, the main decay channels of $K^{*}$ are
$K\pi$ and $K\eta$.
The decay width of $K^{*}$ is derived
\begin{eqnarray}
\lefteqn{\Gamma(q^{2})_{K^{*}}=\frac{f^{2}_{\rho\pi\pi}(q^{2})}
{8\pi}\frac{k^{3}}{\sqrt{q^{2}}m_{K^{*}}}}\nonumber \\
&&+cos^{2}20^{0}
\frac{f^{2}_{\rho\pi\pi}(q^{2})}
{8\pi}\frac{k^{'3}}{\sqrt{q^{2}}m_{K^{*}}},\nonumber \\
&&k=\{{1\over4q^{2}}(q^{2}+m^{2}_{K}-m^{2}_{\pi})^{2}-m^{2}_{K}
\}^{{1\over2}},\nonumber \\
&&k'=\{{1\over4q^{2}}(q^{2}+m^{2}_{K}-m^{2}_{\eta})^{2}-m^{2}_{K}
\}^{{1\over2}}.\nonumber
\end{eqnarray}
In $\Gamma_{K_{1}}(q^{2})$ the
decay modes $K^{*}\pi$, $K\rho$ and
$K\omega$ are included
\begin{eqnarray}
\lefteqn{\Gamma(q^{2})_{K_{1}}=\frac{k}{32\pi}\frac{1}{\sqrt{q^{2}}
m_{K_{1}}}\{(3+{k^{2}\over m^{2}_{K^{*}}})A^{2}(q^{2})_{K^{*}}
}\nonumber \\
&&-A(q^{2})_{K^{*}}B(q^{2}+m^{2}_{K^{*}})\frac{k^{2}}{m^{2}_{K^{*}}}
+{q^{2}\over m^{2}_{K^{*}}}k^{4}B^{2}\}\nonumber \\
&&+{4\over3}\frac{k'}{32\pi}\frac{1}{\sqrt{q^{2}}
m_{K_{1}}}\{(3+{k^{2}\over m^{2}_{\rho}})A^{2}(q^{2})\nonumber \\
&&-A(q^{2})B(q^{2}+m^{2}_{\rho})\frac{k^{2}}{m^{2}_{\rho}}
+{q^{2}\over m^{2}_{\rho}}k^{4}B^{2}\},\nonumber \\
&&k=\{{1\over4q^{2}}(q^{2}+m^{2}_{K^{*}}-m^{2}_{\pi})^{2}
-m^{2}_{K^{*}}
\}^{{1\over2}},\nonumber \\
&&k'=\{{1\over4q^{2}}(q^{2}+m^{2}_{K}-m^{2}_{\rho})^{2}
-m^{2}_{K}\}^{{1\over2}}.
\end{eqnarray}
For $K_{1}(1270)$, \(\Gamma(K_{1}(1270)\rightarrow K^{*}_{0}(1430)\pi)
=25.2MeV\) is included.
The branching ratio
is calculated
\[B(\tau^{-}\rightarrow \bar{K}^{*0}\pi^{-}\nu)=0.23\%,\]
The contribution of the vector current is about $7.4\%$. Therefore,
$K_{a}$ is dominant in this decay.
The data are\\
\(0.38\pm0.11\pm0.13\%\)(CLEO)\\
\(0.25\pm0.10\pm0.05\%\)(ARGUS)\\
There is another decay channel $\tau^{-}\rightarrow K^{*-}\pi^{0}\nu$
whose branching ratio is one half of $B(\tau^{-}\rightarrow\bar{K}^{*0}
\pi^{-}\nu$. The total branching ratio is
\[B(\tau^{-}\rightarrow \bar{K}\pi\nu)=0.35\%,\]
The width is about 230MeV.
\section{$\tau\rightarrow K\rho\nu$ and $K\omega\nu$}
It is the same as $\tau\rightarrow K^{*}\pi\nu$, $K_{a}$
dominates the decay $\tau\rightarrow K\rho\nu$.
Both the vector and axial-vector currents contribute to this decay
mode.
The matrix element of the vector current, $<\bar{K^{0}}\rho^{-}
|\bar{\psi}\lambda_{+}\gamma_{\mu}\psi|0>$
is determined by the abnormal
vertex ${\cal L}^{K^{*}K\rho}$(47).
The axial-vector matrix element $<\bar{K^{0}}\rho^{-}|\bar{\psi}
\lambda_{+}\gamma_{\mu}\gamma_{5}\psi|0>$ is obtained by
substituting
\[K^{*}\rightarrow \rho,\;\;\;K\rightarrow\pi\]
in Eq.(8). Using the same substitutions in Eq.(15),
the distribution of
the decay rate of $\tau\rightarrow K\rho\nu$ is found.
The branching ratio of $\tau\rightarrow K\rho\nu$(two modes
$\bar{K}^{0}\rho^{-}$ and $K^{-}\rho^{0}$)
is computed to be
\begin{equation}
B=0.75\times10^{-3}.
\end{equation}
It is about $18\%$ of $\tau\rightarrow K\pi\pi\nu$.
The vector current makes $8\%$ contribution.
The DELPHI has reported that $\tau\rightarrow K^{*}\pi\nu$ is
dominant the decay $\tau\rightarrow K\pi\pi\nu$ and $K\rho\nu$ decay
mode has not been observed. The recent
ALEPH's results are the $K^{*}\pi$
dominance and a branching ratio of $30\pm 11\%$ for the $K\rho$ mode.

Due to the $SU(3)$ coefficient we expect
\[B(\tau\rightarrow K\omega\nu)={1\over3}B(\tau\rightarrow K\rho\nu).\]
\section{$\tau\rightarrow K^{*}\eta$}
There are vector and axial-vector parts
in this decay. The
calculation of the decay rate is similar to the decay of $\tau
\rightarrow K^{*}\pi\nu$. The vertices ${\cal L}^{K^{*}\bar{K}^{*}
\eta}$ and ${\cal L}^{WK^{*}\eta}$ via the Lagrangian $L^{Vs}$
contribute to the vector part and the vertices ${\cal L}^{K_{1}
K^{*}\eta}$, ${\cal L}^{KK^{*}\eta}$, and ${\cal L}^{WK^{*}\eta}$
via $L^{As}$ take the responsibility for the axial-vector part.
The vertex ${\cal L}^{K^{*}\bar{K}^{*}\eta}$ comes from anomaly.
Using the same method deriving the vertices $\eta vv$ (in Ref.[2])
, it is found
\begin{eqnarray}
\lefteqn{{\cal L}^{K^{*}\bar{K^{*}}\eta}=
-\frac{3a}{2\pi^{2}g^{2}f_{\pi}}
d_{ab8}\varepsilon^{\mu\nu\alpha\beta}\eta\partial_{\mu}K^{a}_{\nu}
\partial_{\alpha}K^{b}_{\beta}}\nonumber \\
&&-\frac{3b}{2\pi^{2}g^{2}f_{\pi}}
\varepsilon^{\mu\nu\alpha\beta}\eta\partial_{\mu}K^{a}_{\nu}
\partial_{\alpha}K^{a}_{\beta},
\end{eqnarray}
where a and b are the octet and singlet component of $\eta$
respectively,
\(a=cos\theta\), \(b=\sqrt{{2\over3}}cos\theta\),
and \(\theta=-20^{0}\). Due to the cancelation
between the two components
the vector matrix element is very small and can be ignored.

The vertices ${\cal L}^{K_{1}K^{*}\eta}$ and ${\cal L}^{KK^{*}\eta}$
contribute to the axial-vector
matrix element and they are derived from the effective Lagrangian
presented in Ref.[2].
\[{\cal L}^{K_{1}K^{*}\eta}=af_{ab8}\{A(q^{2})_{K^{*}}
K^{a}_{\mu}
K^{b\nu}\eta\]
\[-BK^{a}_{\mu}K^{b}_{\nu}\partial^{\mu\nu}\eta\},\]
\[{\cal L}^{K^{*}K\eta}=af_{K^{*}K\eta}f_{ab8}K^{a}_{\mu}(K^{b}
\partial^{\mu}\eta-\eta\partial^{\mu}K^{b}), \]
where $f_{K^{*}K\eta}$ is the same as $f_{\rho\pi\pi}$ in the limit
of \(m_{q}=0\).
The decay width is similar to the one of $\tau\rightarrow K^{*}\pi\nu$
\begin{eqnarray}
\lefteqn{{d\Gamma\over dq^{2}}(\tau^{-}\rightarrow
K^{*-}\eta\nu)=}\nonumber \\
&&{G^{2}\over
(2\pi)^{3}}cos^{2}20^{0}\frac{sin^{2}\theta_{C}}{64m^{3}_{\tau}q^{4}}
(m^{2}_{\tau}-q^{2})^{2}\nonumber \\
&&(m^{2}_{\tau}+2q^{2})
\{(q^{2}+m^{2}_{K^{*}}
-m^{2}_{\pi})^{2}-4q^{2}m^{2}_{K^{*}}\}^{{1\over2}}\nonumber \\
&&\{\frac{1}{2\pi^{4}g^{2}f^{2}_{\pi}}\frac{m^{4}_{K^{*}}+q^{2}
\Gamma^{2}_{K^{*}}(q^{2})}{(q^{2}-m^{2}_{K^{*}})^{2}+q^{2}\Gamma^{2}
_{K^{*}}(q^{2})}\}\nonumber \\
&&[(p\cdot q)^{2}-q^{2}m^{2}_{K^{*}}]\nonumber \\
&&+{3\over4}\{|A|^{2}[1+{1\over12m^{2}_{K^{*}}q^{2}}
(q^{2}-m^{2}_{K^{*}})^{2}]\nonumber \\
&&-(BA^{*}+B^{*}A)\frac{(q^{2}+m^{2}_{K^{*}})}{24m^{2}_{K^{*}}
q^{2}}(q^{2}-m^{2}
_{K^{*}})^{2}\nonumber \\
&&+\frac{|B|^{2}}{48m^{2}_{K^{*}}q^{2}}
(q^{2}-m^{2}_{K^{*}})
^{4}]\}\}.
\end{eqnarray}
The branching ratio is computed
to be
\[B=1.05\times10^{-4}.\]
The contribution of the vector current is $3.3\%$ and
the axial-vector current is dominant.
\section{Strong anomaly of PCAC}
{\bf $\tau\rightarrow\omega\rho\nu$}\\
Only the axial-vector currents contribute to
$\tau\rightarrow\omega\rho\nu$.
At the tree level ${\cal L}^{\omega\rho\pi}$
is the only vertex involved in this decay channel
\begin{equation}
{\cal L}^{\omega\rho\pi}=-\frac{N_{C}}{\pi^{2}g^{2}f_{\pi}}
\varepsilon^{\mu\nu\alpha\beta}\partial_{\mu}\omega_{\nu}\partial
_{\alpha}\rho^{i}_{\beta}\pi^{i}.
\end{equation}
This vertex has been tested by $\tau\rightarrow\omega\pi\nu$,
$\pi^{0}\rightarrow2\gamma$, $\omega\rightarrow\pi\gamma$,
$\rho\rightarrow\pi\gamma$, and $\omega\rightarrow3\pi$.

The matrix element of the axial-vector
current is obtained
\begin{eqnarray}
\lefteqn{
<\omega\rho^{-}|\bar{\psi}\tau_{+}\gamma_{\mu}\gamma_{5}\psi|0>=
-\frac{i}{\sqrt{4E_{1}E_{2}}}
\frac{N_{C}}{\pi^{2}g^{2}}\frac{q_{\mu}}{q^{2}}
\varepsilon^{\lambda\nu\alpha\beta}}\nonumber \\
&&p_{1\lambda}p_{2\nu}\epsilon^{*}_{
\alpha}(p_{1})\epsilon^{*}_{\beta}(p_{2}).
\end{eqnarray}
In the limit \(m_{q}=0\), the axial-vector current is not conserved
in this process.
The pion exchange dominates
this decay.

The decay width is
\begin{eqnarray}
\lefteqn{\Gamma=\frac{G^{2}}{128m_{\tau}}
\frac{cos^{2}\theta_{C}}{(2\pi)^{3}}\frac{9}{\pi^{4}g^{4}}\int^{m^{2}_
{\tau}}_{q^{2}_{min}}dq^{2}
{1\over q^{6}}(m^{2}_{\tau}-q^{2})^{2}}\nonumber \\
&&\int^{(\sqrt{q^{2}}-m_{\omega})^{2}}_{4m^{2}_{\pi}}dk^{2}
[(q^{2}+m^{2}_{\omega}-k^{2}
)^{2}-4q^{2}m^{2}_{\omega}]^{{3\over2}}\nonumber \\
&&{1\over\pi}
\frac{\sqrt{k^{2}}\Gamma_{\rho}(k^{2})}{(k^{2}-m^{2}_{\rho})^{2}+k^{2}
\Gamma^{2}_{\rho}(k^{2})},
\end{eqnarray}
where \(q^{2}_{min}=2m_{\pi}m_{\tau}+\frac{m_{\tau}m^{2}_{\omega}}{
m_{\tau}-2m_{\pi}}\),
$k^{2}$ is the invariant mass of the two pions and
\begin{eqnarray}
\Gamma_{\rho}(k^{2})=\frac{f^{2}_{\rho\pi\pi}(k^{2})
}{48\pi}{k^{2}\over
m_{\rho}}(1-4{m^{2}_{\pi}\over k^{2}})^{3}.
\end{eqnarray}
The branching ratio is computed to be
\[B=0.16\times10^{-4}.\]

The nonconservation of the quark axial-vector currents
found in the matrix
elements shows that there is anomaly in the PCAC.
Taking one abnormal term as an example of the strong anomaly of
the PCAC, the PCAC with strong anomaly is written as
\begin{equation}
\partial^{\mu}\bar{\psi}\tau_{i}\gamma_{\mu}\gamma_{5}\psi=
-m^{2}_{\pi}f_{\pi}\pi_{i}+\frac{N_{C}}{\pi^{2}g^{2}}
\varepsilon^{\mu\nu\alpha\beta}\partial_{\mu}\omega_{\nu}
\partial_{\alpha}\rho^{i}_{\beta}.
\end{equation}
The abnormal term
originates in the WZW anomaly.

On the other hand, using the VMD we derive
\begin{eqnarray}
\lefteqn{\partial^{\mu}\bar{\psi}\tau_{i}\gamma_{\mu}\gamma_{5}\psi=
-m^{2}_{\pi}f_{\pi}\pi_{i}+\frac{N_{C}}{\pi^{2}g^{2}}
\varepsilon^{\mu\nu\alpha\beta}\partial_{\mu}\omega_{\nu}
\partial_{\alpha}\rho^{i}_{\beta}}\nonumber \\
&&+\frac{\alpha}{4\pi}\varepsilon^{\mu\nu\alpha\beta}F_{\mu\nu}
F_{\alpha\beta}\delta_{3i}\nonumber \\
&&+\frac{e}{4\pi^{2}g}\varepsilon^{\mu\nu\alpha\beta}F_{\mu\nu}
\partial_{\alpha}\rho^{i}_{\beta}\nonumber \\
&&+\frac{3e}{4\pi^{2}g}
\varepsilon^{\mu\nu\alpha\beta}\partial_{\mu}\omega_{\nu}F_{\alpha
\beta}\delta_{3i}.
\end{eqnarray}

The background of this process comes from
$\tau\rightarrow\omega(\pi\pi)_{non\rho}\nu$.
Only the axial-vector currents contribute to
$\tau\rightarrow\omega(\pi\pi)_{non\rho}\nu$.
There are two kinds of vertices
involved in this decay channel. These vertices are derived from
the effective chiral theory of mesons[2] in the chiral limit.
\begin{enumerate}
\item The vertices $\omega\pi\pi\pi$ and $\omega a_{1}\pi\pi$
are
\begin{eqnarray}
\lefteqn{{\cal L}^{\omega\pi\pi\pi}
=\frac{2}{\pi^{2}gf^{3}_{\pi}}(1-{6c\over g}+{6c^{2}
\over
g^{2}})\varepsilon^{\mu\nu\alpha\beta}}\nonumber \\
&&\epsilon_{ijk}\omega_{\mu}
\partial_{\nu}\pi_{i}\partial_{\alpha}\pi_{j}\partial_{\beta}\pi_{k}
\nonumber \\
&&{\cal L}^{\omega a_{1}\pi\pi}=
-\frac{6}{\pi^{2}g^{2}f^{2}_{\pi}}(1-{1\over2\pi^{2}g^{2}})
^{-{1\over2}}\nonumber \\
&&(1-{2c\over g})\varepsilon^{\mu\nu\alpha\beta}
\epsilon_{ijk}\partial_{\nu}\omega_{\mu}a^{i}_{\alpha}
\partial_{\beta}\pi_{k}.
\end{eqnarray}
These two vertices are from the WZW anomaly[2].
Using these vertices and ${\cal L}^{A}$,
the matrix element of the axial-vector current is obtained
\begin{eqnarray}
\lefteqn{<\omega\pi^{0}\pi^{-}|\bar{\psi}\tau_{+}\gamma_{\mu}
\gamma_{5}\psi|0>^{(1)}
=\frac{i}{\sqrt{8\omega_{1}\omega_{2}E}}}\nonumber \\
&&\frac{6}
{\pi^{2}gf^{2}_{\pi}}\varepsilon^{\nu\lambda\alpha\beta}\epsilon^{*}_
{\lambda}p_{\alpha}(k_{2}-k_{1})_{\beta}\nonumber \\
&&\{(\frac{q_{\mu}q_{\nu}}{q^{2}}-g_{\mu\nu})(1-{2c\over g})\nonumber \\
&&\frac{g^{2}f^{2}_{a}m^{2}_{\rho}-i\sqrt{q^{2}}\Gamma_{a}(q^{2})}{
q^{2}-m^{2}_{a}+i\sqrt{q^{2}}\Gamma_{a}(q^{2})}\nonumber \\
&&+\frac{q_{\mu}q_{\nu}}{q^{2}}{2c\over g}
(1-{2c\over g})\},
\end{eqnarray}
where $k_{1}$, $k_{2}$, and p are momentum of $\pi^{0}$, $\pi^{-}$, and
$\omega$ respectively, \(q=k_{1}+k_{2}+p\).
The equation
shows that the matrix element obtained from the WZW anomaly is
not
conserved in the limit of \(m_{q}=0\). The pion exchange is dominant
in the term which violates the conservation of the quark axial-vector
current in the chiral limit.
\item The second kind of vertices are
vertices $a_{1}\rho\pi$, $\rho\pi\pi$, $W\rho\pi$
and $\omega\rho\pi$.
The vertex $\omega\rho\pi$ is from the WZW anomaly.
Using the vertex ${\cal L}^{a_{1}\rho\pi}$, the decay width of
$a_{1}$ meson is derived
\begin{eqnarray}
\lefteqn{\Gamma_{a}(q^{2})
=\frac{k}{12\pi m_{a}\sqrt{q^{2}}}\{(3+{k^{2}\over
m^{2}_{\rho}})A^{2}}\nonumber \\
&&-\frac{k^{2}}{m^{2}_{\rho}}
(q^{2}+m^{2}_{\rho})AB
+\frac{q^{2}}{m^{2}_{\rho}}k^{4}
B^{2}\},
\end{eqnarray}
where
\[k=\{{1\over4q^{2}}(q^{2}+m^{2}_{\rho}-m^{2}_{\pi})^{2}-m^{2}_{\rho}\}
^{{1\over2}}.\]
Using ${\cal L}^{A}$
and the vertices,
the second part of the matrix element of the
axial-vector current is obtained
\begin{eqnarray}
\lefteqn{<\omega\pi^{0}\pi^{-}|\bar{\psi}\tau_{+}\gamma_{\mu}\gamma_{5}
\psi|0>^{(2)}}\nonumber \\
&&=\frac{i}{\sqrt{8\omega_{1}\omega_{2}E}}(\frac{q_{\mu}q_{\nu}}
{q^{2}}-g_{\mu\nu})\nonumber \\
&&\frac{3}{\pi^{2}g^{2}
f_{\pi}}\frac{g^{2}f_{a}m^{2}
_{\rho}-if^{-1}_{a}
q\Gamma_{a}(q^{2})}{q^{2}-m^{2}_{a}+iq\Gamma_{a}(q^{2})}
\nonumber \\
&&\varepsilon^{\nu\lambda\alpha\beta}\epsilon^{*}_{\lambda}p_{\alpha}
\{
\frac{A(k^{2})k_{2\beta}}{k^{2}-m^{2}_{\rho}+i\sqrt{k^{2}}
\Gamma_{\rho}(k^{2})}\nonumber \\
&&-\frac{A(k^{'2})k_{1\beta}}{k^{'2}-m^{2}_{\rho}+i
\sqrt{k^{'2}}\Gamma_{\rho}(k^{'2})}\}\nonumber \\
&&+\frac{i}{\sqrt{8\omega_{1}\omega_{2}E}}
(\frac{q_{\mu}q_{\nu}}
{q^{2}}-g_{\mu\nu})\frac{3}{\pi^{2}g^{2}
f_{\pi}}\nonumber \\
&&\frac{g^{2}f_{a}m^{2}
_{\rho}-iqf^{-1}_{a}\Gamma_{a}(q^{2})
}{q^{2}-m^{2}_{a}+iq\Gamma_{a}(q^{2})} \nonumber \\
&&\varepsilon^{\sigma\lambda\alpha\beta}
\epsilon^{*}_{\sigma}p_{\lambda}k_{2\alpha}k_{1\beta}(-B)\nonumber \\
&&\{\frac{k_{2\nu}}{k^{'2}-m^{2}_{\rho}+i\sqrt{k^{'2}}\Gamma_{\rho}
(k^{'2})}\nonumber \\
&&+\frac{k_{1\nu}}{k^{2}-m^{2}_{\rho}+i\sqrt{k^{2}}\Gamma_{\rho}
(k^{2})}\},
\end{eqnarray}
where $k_{1}$, $k_{2}$, p are momentum of $\pi^{0}$, $\pi^{-}$, and
$\omega$ mesons respectively,
\(q=p+k_{1}+k_{2}\), \(k=q-k_{1}\), \(k^{'}=q-k_{2}\),
$A(k^{2})$ and $A(k^{'2})$ are defined by Eq.(16) by taking
\(p^{2}=k^{2}, k^{'2}\) respectively.
This part of the matrix element observes the axial-vector current
conservation in the limit \(m_{q}=0\) and there is $a_{1}$ dominance.
\end{enumerate}
The theory makes explicit prediction
on this decay
The branching ratio is computed to be
\[B=0.37\%.\]
The data are
$0.39\pm0.04\pm0.04\%$
(CLEO) and
$0.41\pm0.08\pm0.06\%$(ALEPH).
\section{Summary of other results}
The decay $\tau\rightarrow K^{*}K\nu$ is calculated[3]. Both the vector
and axial-vector currents contribute to this process. For the vector
part only the $\rho$ meson is taken into account. The vertex is
$K^{*}K\rho$ from WZW anomaly. The vector current makes $92.5\%$
contribution. The peak of the spectrum is at 1.5GeV. The theoretical
result of the decay rate is compatible with the data.

The $K^{*}$ meson is dominant the decay $\tau\rightarrow\eta K\nu$[3].
The vertex $\eta K^{*}K$ has a form factor which is the same as
$f_{\rho\pi\pi}(q^{2})$. We obtain[3] \(B=2.22\times 10^{-4}\).

The decays, $\tau\rightarrow\eta\pi\pi\nu$ and $\eta'\pi\pi\nu$ are
calculated in terms the theory[3]. Besides the vertices $\eta\rho\rho$
and $\eta'\rho\rho$, there are contact terms. It is found that the
contributions of the contact terms are negligible. The theoretical
results are
\[B(\tau\rightarrow\eta2\pi\nu)=1.9\times10^{-3},\]
\[B(\tau\rightarrow\eta`2\pi\nu)=0.44\times10^{-5}.\]

\section{Conclusions}
\begin{enumerate}
\item So far, the theory agrees with data reasonably well.
\item In two flavor cases the $a_{1}$ dominance is associated with
PCAC.
\item The dominance of pion exchange is associated with the strong
anomaly of PCAC.
\item $K_{a}$ dominates the axial-vector currents of \(\Delta s=1\).
\item The momentum dependeces of the vertices VPP and AVP are very important in
understanding the $\tau$ mesonic decays.
\item The new formula of the resonance of the axial-vector meson doesn't have
the ``chiral limit`` at low $q^{2}$.
\end{enumerate}
\section*{ACKNOWLEDGEMENTS}
The author would like to thank E.Braaten, J.Smith, and R.Stroynowsky
for discussion.

\end{document}